\documentclass[a4paper,12pt]{article}
\usepackage[cp1251]{inputenc}
\usepackage[russian]{babel}

\begin{document}

{\Large\bf Lema$\hat{\imath}$tre and Hubble: What was discovered
-- if any -- in 1927-29?}

\vspace{1cm}

A.D. Chernin

\vspace{0.5cm}

Sternberg Astronomical Institute, Moscow University

\vspace{1cm}

{\bf

The Big Bang predicted theoretically by Friedmann could not be
discovered in the 1920th, since global cosmological distances
(more than 300-1000 Mpc) were not available for observations at
that time. In 1927-29, Lema$\bf\hat{\imath}$tre and Hubble studied
receding motions of galaxies at local distances of less than 20-30
Mpc and found that the motions followed the (nearly) linear
velocity-distance relation, known now as Hubble's law. For
decades, the real nature of this phenomenon has remained a
mystery, in Sandage's words. After the discovery of dark energy,
it was suggested that the dynamics of local expansion flows is
dominated by omnipresent dark energy, and it is the dark energy
antigravity that is able to introduce the linear velocity-distance
relation to the flows. It implies that Hubble's law observed at
local distances was in fact the first observational manifestation
of dark energy. If this is the case, the commonly accepted
criteria of scientific discovery lead to the conclusion: In 1927,
Lema$\bf\hat{\imath}$tre discovered dark energy and Hubble
confirmed this in 1929. }

\vspace{1cm}

\section{Introduction}

It has seemingly been taken for granted that in 1929 Hubble
discovered exactly what Friedmann predicted several years before,
in 1922 (see, for example, "The brief history of time" by Stephen
Hawking [1], a book issued in a number of copies which is much
larger than the total number of copies of all other books on
cosmology ever published). A book by Alexander Sharov and Igor
Novikov declares in its title: "Edwin Hubble, the Discoverer of
the Big Bang Universe" [2]. Einstein was among the first
physicists and astronomers who adopted or shared this view (but
not Friedmann who died in 1925).

A non-traditional point was however made by Steven Weinberg in
"The First Three Minutes"[3]: "Actually, a look at Hubble's data
leaves me perplexed how he could reach such a conclusion --
galactic velocities seem almost uncorrelated. In fact, we would
not expect any neat relation of proportionality between velocity
and distance for these 18 galaxies -- they are all much too close,
none been further than the Virgo Cluster. It is difficult to avoid
the conclusion that... Hubble knew the answer he wanted to get."

In [1-3], Lema$\hat{\imath}$tre is not given any credit for the
linear velocity-distance relation (the very his name cannot be
found in [1]), though the discovery of the relation is explicitly
reported in his 1927 paper long known to many cosmologists. The
observational data he used (especially as presented in a
velocity-distance diagram [4]) look somewhat more scattered than
that in Hubble's diagram of 1929. So Weinberg's challenge extends
equally to Lema$\hat{\imath}$tre's result as well.

Contrary to Weinberg, Alan Sandage, who was Hubble's successor in
observational cosmology on Mount Wilson and Mount Palomar,
accepted Hubble's law for local galaxies as an important
well-established empirical fact. He had however doubts about its
cosmological interpretation [5-7].

The recent debate on the history of Hubble's law [4,8-11] revives
interest to earlier controversial views [1-3,5-7] and rises again
a question on the essence of the matter: What was actually
discovered -- if any -- in 1927-29?

\section{Paradox}

Sandage's argument [5-7] of 1972-99, was, in brief, as follows.
Friedmann's cosmology describes a Universe which has the uniform
(homogeneous) matter distribution and expands in accordance with
the linear velocity-distance relation. The linearity and the
uniformity are linked: the matter distribution may be uniform and
preserve its uniformity, if and only if the expansion velocity is
directly proportional to the distance at any moment of time. The
observed Universe is indeed uniform on average over spatial scales
of more than 300-1000 Mpc which is the size of the cosmic cell of
uniformity. Friedmann's cosmology is applicable to such large
distances only, and it says nothing about local spatial scales of
less than 2-3 Mpc (or 20-30 Mpc, as it became clear after the
distance revision made by Sandage in the late 1950th).

Does it mean that Hubble's law has nothing in common with global
cosmological expansion?

Note that not only in cosmology, but also in dynamics of
gravitating medium in general, the linear velocity-distance
relation assumes uniformity of matter distribution and vice verse.
Therefore one should expect matter uniformity in the area where
the velocity-distance linearity is observed. But the real matter
distribution is highly non-uniform on local scales of a few dozen
Mpc.

Sandage introduced the notion of "expansion flows" for systems of
receding  galaxies with nearly linear velocity-distance relation
to describe the observational situation in most obvious way. When
deviations from linearity (velocity dispersion) are small, the
flow is considered "quiet". The quietness of expansion flows has
repeatedly been confirmed in increasingly precise observations
both inside and outside the cosmic cell of uniformity. In 1999 [7]
and again in 2006 [12], Sandage (together with his collaborators)
reported that expansion flows were quiet in the distance interval
from a few Mpc to the global cosmological scales. It is especially
puzzling that the rate of expansion (the velocity-to-distance
ratio known as Hubble's factor in cosmology) is the same within
10-15\% accuracy for all these distances and for the global
expansion as well.

How may this be possible, if local flows have nothing in common
with cosmology?

Thus, Sandage pointed out a paradox: 1) Linear flows are observed
inside the cosmic cell of uniformity where the matter distribution
is highly non-uniform and such flows cannot exist. 2) The
expansion rates of the local flows are practically the same as in
the Universe as a whole.

In 1999, 70 years after Hubble, Sandage concluded: "We are still
left with the mystery" [7]. A year later, in 2000, a solution to
the paradox was suggested [13]: 1) Local flows of expansion are
nearly linear due to dark energy. 2) Omnipresent dark energy
dominates the dynamics of both local and global flows and makes
their expansion rates be (almost) identical.

Sandage and his colleagues commented this suggestion in 2006: "No
viable alternative to vacuum [dark] energy is known at present.
The quietness of the Hubble flow lends support for the existence
of vacuum energy" [12].

\section{Dark energy in quiet flows}

Dark energy was discovered in 1998-99 at largest horizon-scale
cosmological distances [14,15]. It contributes about 3/4 to the
total mass/energy balance of the observed Universe as a whole. Its
microscopic structure is completely unknown -- this is considered
as the most severe problem of fundamental physics and astronomy of
the 21st century.

In macroscopic description, dark energy may adequately be treated
as a vacuum-like perfectly uniform fluid which produces
antigravity. The observed effective (anti)gravitating density of
dark energy is 6 times the gravitating mean cosmic matter (dark
matter and baryons) density for the Universe as a whole at the
present epoch. Because of this dark energy controls (mainly) the
observed global cosmological expansion and makes the Universe
expand with acceleration. These are the key features of the
currently standard $\Lambda$CDM cosmological model in which dark
energy is represented by Einstein's cosmological constant
$\Lambda$.

In [13] (see also [16-22]), it was demonstrated that dark energy
could act and even dominate not only globally, but also locally,
inside the cosmic cell of uniformity. Since dark energy exists
everywhere and has a perfectly the same density in any point of
space, it makes the whole world more uniform on local scales. The
effect is strong in low-density areas outside large matter
overdensities such as groups and clusters of galaxies. Local
expansion flows are observed just in such low-density areas. In
these areas, flow galaxies move (almost) as "test particles" on
the dynamical background dominated by dark energy.

In terms of hydrodynamics, if dark energy dominates dynamics of
expansion flows, it brakes the hard link between the kinematics of
the flows and the matter distribution in them and around. Because
of this, a highly non-uniform matter distribution becomes
compatible with the quietness of an expansion flow. In this way,
the first aspect of the paradox above is eliminated. The second
aspect is also resolved since dark energy dominates both local and
global flows. The local and global flows do not "know" about each
other, but the both are (mainly) controlled by the same physical
agent which is omnipresent perfectly uniform dark energy.

Quantitatively, when dark energy dominates, the expansion rate
(both local and global) is determined (mainly) by dark energy
only; as a result, the rate should be close to the universal value
$H_{\Lambda} = (\frac{8\pi}{3}\rho_{\Lambda})^{1/2} \simeq 60$
km/sec/Mpc, where $\rho_{\Lambda}$ is the density of dark energy
measured in global observations [14,15]. According to the
$\Lambda$CDM model, the observed global expansion rate is $ H_0 =
70-72$ km/s/Mpc. The local rate found recently by Sandage and
collaborators [12] is $H = 63$ km/s/Mpc for the distance interval
from 4 to 200 Mpc. Both rates (measured with the accuracy of about
10\%) are indeed close to the universal value $H_{\Lambda}$, and
because of this they prove to be close to each other.

Interestingly enough, the relation above together with the
observed local expansion rate $H$ might be used to estimate
(approximately) the dark energy density: $\rho_{\Lambda} \sim
\frac{3}{8\pi}H^2$. With the local figure for $H$ from [12], the
dark energy density estimated in this way is very close (if not
exactly equal) to the value measured at the horizon-scale
distances.  Note that such an approximate evaluation of the dark
energy density might be made even in 1927-29 with the use of the
observed expansion rates (though overestimated at that time).

\section{The very local flow}

Karachentsev and his collaborators [23] have recently found (with
the use of the HST) a nearly linear velocity-distance relation in
the "very local" flow of dwarf galaxies at distances less than 3
Mpc from us. Well studied both observationally and theoretically
(see [24] and references therein), the flow presents a good (and
most close to us) example of the general picture sketched briefly
above. Recall that our Galaxy, the Milky Way, together with the
giant Andromeda Nebula and dozens less massive galaxies form the
Local Group of galaxies. This is a gravitationally bound
quasi-stationary system of 2 Mpc across embedded -- as all bodies
of nature -- in the uniform dark energy background. Around the
group, at distances 1-3 Mpc from the group barycenter, two dozen
dwarf galaxies move away from the group forming the very local
expansion flow. The flow is quiet: it follows closely to the
linear velocity-distance relation [23].

The force field in the flow area is a sum of the gravity produced
by the matter (dark matter and baryons) of the group and the
antigravity produced by the dark energy background. The
selfgravity of the flow dwarfs contributes little to the force
field, and they may reasonably be considered as test particles.
Estimates indicate that the dark energy antigravity is stronger
than the gravity of the Local Group at distances larger than
$\simeq 1$ Mpc from the group barycenter. Since antigravity
dominates in the flow area, the flow is accelerated by dark
energy. Our models show also that the accelerated flow might be
rather chaotic initially, but it is getting more and more regular
and quiet with time under the action of the dark energy
antigravity. Asymptotically, the flow becomes exactly linear, and
its expansion rate approaches the universal value $H_{\Lambda}
\simeq 60$ km/sec/Mpc which is determined by the dark energy
density only (see Sec.2). A similar asymptotic behavior is
prescribed by the $\Lambda$CDM model to the global cosmological
expansion: its expansion rate (the Hubble factor) tends to the
same universal value $H_{\Lambda}$. Since the observed values of
$H$ and $H_0$ are close to $H_{\Lambda}$, the present-day states
of both local and global flows are not far from their common
asymptotic state -- as it should be indeed where and when dark
energy dominates.

Quiet local flows of expansion have also been observed recently
around two other nearby groups of galaxies and around the Virgo
Cluster of galaxies. The nearly linear velocity-distance relation
and similar expansion rates are characteristic for each of them.
Dark energy domination is recognized to control the major trend of
the flow evolution and determines the asymptotic value of the
expansion rate, while small individual deviations from this are
due to specific local conditions in the flows and around them. The
structure, dynamics and the very origin of the local flows are due
to the local gravity-antigravity interplay and little affected by
the global Big Bang (not saying about "initial inflation").

\section{Conclusions}

To summarize, neither Lema$\hat{\imath}$tre, nor Hubble discovered
the Big Bang in 1927-29. This has been done decades after that by
Sandage and other astronomers (including the authors of the works
[14,15]) who have extended extragalactic observations to the truly
cosmological distances of 300-1000 Mpc and more. At local
distances of less than 20-30 Mpc, Lema$\hat{\imath}$tre in 1927
and then Hubble in 1929 dealt with accidental sets of
galaxy-members of local expansion flows. These galaxies preserve
in their quiet collective kinematics a dynamical signature of dark
energy. The signature is the (nearly) linear velocity-distance
relation. Recognized empirically at local distances, this relation
has occurred to be the first observational manifestation of
omnipresent dark energy.

Thus, the present-day understanding of the essence of the matter
and commonly accepted criteria of scientific discovery lead to the
conclusion: In 1927, Lema$\hat{\imath}$tre discovered dark energy
and Hubble confirmed this in 1929.

\vspace{0.3cm}

I thank G. Byrd, Yu. Efremov, I. Karachentsev, D. Makarov, P.
Teerikorpi, M. Valtonen for collaboration and many useful
discussions.

\section*{References}

\hspace{0.5cm} 1. Hawking S. "A brief history of time" Bantam Dell
Pub. Group, 1988

2. Sharov A.S., Novikov I.D. "Edvin Hubble, the Discoverer of the
Big Bang Universe" Cambridge: Cambridge University Press, 1993

3. Weinberg S. "The first three minutes" Basic Books, 1993

4. Block D.L. arXiv:1106.3928, 2011

5. Sandage A. et al.   Astrophys. J. 172, 253, 1972

6. Sandage A.  Astrophys. J. 307, 1, 1986

7. Sandage A.   Astrophys. J. 527, 479, 1999

8. Nussbaumer H, Bieri L. arXive:1107.2281, 2011

9. van den Bergh S. arXiv:1106.1195, 2011

10. Reich E.S. Nature, 27 June, 2011

11. Luminet J.-P. arXive:1105.6271, 2011

12. Sandage A. et al.  Astrophys. J. 653, 843, 2006

13. Chernin A., Teerikorpi P., Baryshev Yu. arXive:0012021; Adv.
Space Res. {\bf 31}, 459 (2003)

14. Perlmuter S., Aldering G., Goldhaber G. et al.  ApJ, 517, 565,
1999

15. Riess A.G., Filippenko A.V., Challis P. et al.  AJ, 116, 1009,
1998

16. Chernin A.D. Physics-Uspekhi, 44, 1099, 2001

17. Chernin A.D. Physics-Uspekhi 51, 253, 2008

18. Chernin A.D., Karachentsev I.D., Valtonen M.J, et al. A\&A
415, 19, 2004

19. Chernin A.D., Teerikorpi P., Valtonen M.J. et al. A\&A 507,
1271, 2009

20. Teerikorpi P, Chernin, A.D., Karachentsev, I.D., Valtonen M.J.
A\&A 483, 383, 2008

21. Byrd G.G., Chernin A.D. \& Valtonen M.J. Cosmology:
Foundations and Frontiers. Editorial URRS: Moscow, 2007

22. Karachentsev I.D., Chernin A.D., Teerikorpi P. Astrophysics
46, 491, 2003

23. Karachentsev I.D. et al.  MNRAS 393, 1265 (2009)

24. Teerikorpi P., Chernin A.  A\&A 516, 93 (2010)

\end{document}